\documentstyle[aps,prl,psfig,epsfig]{revtex}
\newcommand{\be}{\begin{equation}}
\newcommand{\ee}{\end{equation}}
\newcommand{\bea}{\begin{eqnarray}}
\newcommand{\eea}{\end{eqnarray}}

\begin{document}

\title{ Hysteretic Optimization}
\author{ G. Zar\'and$^{1,2}$, F. P\'azm\'andi$^{2,3,5}$, K. F. P\'al$^{4}$, and 
G. T. Zim\'anyi$^{3}$}
\address{$^{1}$ Lyman Physics laboratory, Harvard University, Cambridge MA 02145}
\address{$^{2}$ Research Group of the Hungarian Academy of Sciences, 
Institute of Physics, TU Budapest, H-1521 Hungary}
\address{$^{3}$ Department of Physics, University of California, Davis, CA 95616}
\address{$^{4}$ Institute of Nuclear Research of the Hung. Acad. of Sci., 
Debrecen, P.O.B 51, H-4001 Hungary}
\address{$^{5}$ Department of Theoretical Physics, University of Debrecen, P.O. 
Box 5, H-4010 Hungary}
\twocolumn[\hsize\textwidth\columnwidth\hsize\csname
@twocolumnfalse\endcsname

\date{\today}

\maketitle

\begin{abstract}
We propose a new optimization method based on a
demagnetization procedure well known in magnetism.
We show how this procedure can be applied as a general
tool to search for optimal solutions in any system where 
the configuration  space is endowed with a suitable `distance'. 
We  test the new algorithm on frustrated magnetic models 
and the traveling salesman problem. 
We find that the new method successfully competes with similar
basic algorithms such as simulated annealing.
\end{abstract}
\pacs{02.60.Pn,75.10.Nr,75.50.Lk}
]
\narrowtext

It has been observed long time ago that disordered  materials 
can be brought into a remarkably stable state through annealing, 
i.e. cooling down the material rather slowly. This simple observation 
inspired  Kirkpatrick, Gelatt, and Vecchi in their pioneering work 
\cite{kirkpatrick} to  investigate how close this annealing 
procedure takes models with glassy  properties  to their  ground state,  
and lead them to the invention of the by now widely used simulated annealing 
(SA) procedure. The SA revealed  the crucial  role the external noise 
can play in optimization:  Thermal noise can help to escape high-energy 
local minima. 
In the present work we investigate another procedure 
that is commonly used  to demagnetize disordered  magnets
and is experimentally known to result in a very stable state: 
The application of an oscillating external  field (see Figure 1).  
This procedure makes use of  another type of noise 
which is typical in magnetic  systems, namely {\em random external 
fields}. As we shall demonstrate below for various models, a simple 
generalization of this zero-temperature `a.c. demagnetization' 
is able to give  systematically better  and better approximations 
to the ground state of these models, 
and is in many cases  5-10-times faster than SA.  We shall show how this 
method  can be applied to  practically {\em any}  disordered model, 
thereby resulting in a new optimization procedure,  that 
we  call  {\em hysteretic  optimization} (HO). 

Finding optimal solutions of complex problems depending on a
large number of parameters is an equally important and difficult
task \cite{papa}. Examples range from integrated circuit design, through
portfolio selection on the stock market \cite{stock} and calculating protein
folding, to teaching artificial neural networks, to name a few.
The simultaneous presence of randomness and frustration is what
makes these problems so hard: Disorder is caused by the non-regular
dependence of the quality of the solution 
on the configuration, and frustration is brought in by the competition of 
mutually exclusive different ``good'' properties.
As a result, on one hand, a naiv search often gets stuck in 
spurious minima while, on the
other hand, comparably good solutions can be found with quite
different configurations.

\begin{figure}[b]
\epsfxsize=2.5in
\begin{center}
\epsfbox{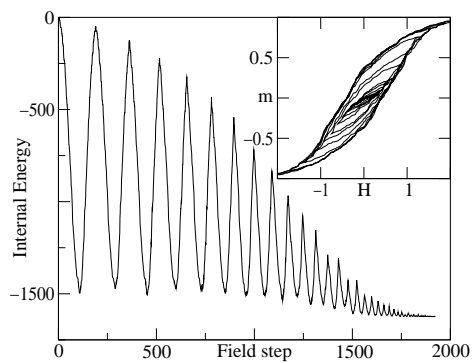}
\end{center}
\vspace*{0.05cm}
\caption{\label{fig:loop} The total
internal energy (without the Zeeman term) of a three 
dimensional ($L=10$) Edwards-Anderson spin glass during 
a.c. demagnetization. The inset shows the corresponding 
magnetization curve.}
\end{figure}

It is
important to note that most hysteretic systems fulfill the above
requirements of complexity. Hysteresis 
implies the presence of many metastable states caused mainly 
by disorder. In the case of magnetic materials, the other important 
ingredient, frustration, is furnished by the magnetostatic 
interaction, which can be ferro- or antiferromagnetic depending
on the relative orientation of the dipoles. 
This analogy between magnetic sytems and
optimization suggests that part of knowledge 
accumulated through decades
by hysteresis research \cite{bertotti,dellatorre} will eventually 
prove useful in optimization (and vice versa).
Indeed, 
a simple but very frequently used  hysteresis model, 
the Preisach model \cite{bertotti,dellatorre,preisach}, can
be put in its ground state by a.c. demagnetization. 

In this letter, we proceed by studying in detail the energetics of
a.c. demagnetization for spin glass models. Then we
show how this simple method can be modified to obtain
a hysteretic optimization  technique, which systematically
approaches the ground state \cite{noproof}, and we 
compare the new algorithm with SA.
We also present how HO can be combined with cluster
renormalization, 
and apply the general formalism of HO for the  
traveling salesman problem (TSP).  

Many optimization problems can be formulated 
in terms of interacting Ising spins  ($\sigma=\pm 1$). 
Therefore we  studied first  two classical Ising spin 
glass models \cite{sk} with a Hamiltonian:
\begin{equation}
{\cal H}=-{1\over 2} \sum_{i,j}^N J_{ij}\sigma_i\sigma_j -
                  H  \sum_i^N\xi_i\sigma_i.
\label{eq:energy}
\end{equation}
In the case of the Sherrington-Kirkpatrick (SK) model
for any $(i,j)$ pair
$J_{ij}=z_{ij}/\sqrt N$, where $z_{ij}$ is a random gaussian number 
with zero mean and unit variance. In the other spin glass we considered, 
the three dimensional Edwards-Anderson (EA) model, 
$J_{ij} =z_{ij}$, but only for the nearest-neighbor couplings. 
The direction $\xi_i=\pm 1$ of  the external field $H$ 
randomly changes  from site to site. 
Due to a spin-gauge symmetry, any choice of $\xi_i$ is equivalent,
as long as it is not correlated with $J_{ij}$. Using this
symmetry, and also the analogy of the magnetic energy from equation
(\ref{eq:energy}), we call $m=1/N\sum\xi_i\sigma_i$ magnetization
throughout the paper.
These sytems are in the focus of research
both in magnetism and in optimization.  In magnetism their dynamics
and the nature of the ordered state is still highly debated 
\cite{young,middleton}. From the optimization point of view,
they are interesting because
they are thought to be so-called NP-hard problems \cite{hartmann}.
The major difference between the two models lies in the range
of interactions. The short-ranged EA model may show some kind
of clusterization (``droplets''), which is very unlikely for the
infinite-ranged SK model.

\begin{figure}
\epsfxsize=2.5in
\begin{center}
\epsfbox{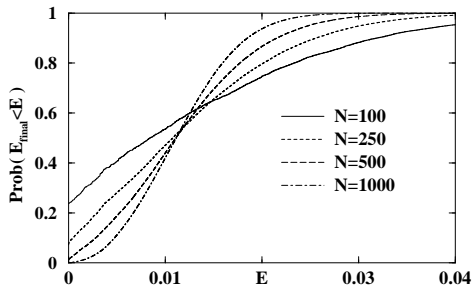}
\end{center}
\vspace*{0.05cm}
\caption{ 
\label{fig:sizedep}
Size dependence of the integrated energy distribution 
of states produced by a.c. demagnetization
for the SK model ($\gamma = 0.9$).
$E$ is the energy per spin measured from the ground state of 
each sample.}
\end{figure}

We tested 
first the energetic properties of a.c. demagnetization on
the SK model  \cite{bertottisk}.  Many properties 
of this model can be calculated exactly \cite{sk},
and it is an ideal test system. We know, e.g., that its ground state energy
is $-0.765/{\rm spin}$   as $N\to\infty$.
Starting from a random initial state,  a simple `quench', 
when randomly chosen unstable spins are aligned one-by-one 
with their local fields,
$h_i=\sum_j J_{ij}\sigma_j+H\xi_i$,
ends up with (locally) stable states of energy around $-0.70$ per 
spin at $H=0$. 

We started  the a.c. demagnetization with an external 
field that saturates the spins,
i.e. an $H=H_1$ large enough that the $\sigma_i=\xi_i$ state
($m=1$) be stable.
Then we decreased the field $H$ in small steps, 
each time making a few spins unstable,
and `quenching' the system to a nearby stable state
using  sequential dynamics.
When $H$ reached the `turning point' $H_2=-\gamma_1 H_1$,
we started to increase the field again until $H_3=-\gamma_2 H_2$,
and turned back again and again at points $H_n=-\gamma_{n-1} H_{n-1}$,
until the amplitude of the loop, $|H_n|$, reached the order
of the field steps.
The resulting ``spiraling'' magnetization curve is shown in the
inset of Figure \ref{fig:loop}.
We performed  a.c. demagnetization
with loops which were in average reduced by a factor of 
$\gamma=\langle \gamma_i\rangle=0.9$ after each
turning point (we used $\gamma_i$-s between $0.8$ and  $1.0$). 
We found that, as expected, lower $\gamma$ gives
poorer results, but a $\gamma$ even closer to one did not make a
considerable  improvement.  However, randomly varying the turning 
points, $\gamma_i$, increased the efficiency.

Figure ~\ref{fig:sizedep}  shows for different system sizes
the probability of finding a configuration close to the ground
state \cite{gs} of the SK model with a.c. demagnetization.
For each $N$ we averaged over several realizations of 
the couplings and ran  a.c. demagnetization
many times with different random field 
directions $\{\xi_i\}$ to generate the distributions presented.
The finite intercept at $E=0$
for smaller systems signals a finite probability of finding 
the ground state, but this probability goes to zero as the 
size increases: for very large systems  a.c. demagnetization
typically gives a state with  an energy   $0.012\times N$ above 
the ground state.

\begin{figure}
\epsfxsize=2.6in
\begin{center}
\epsfbox{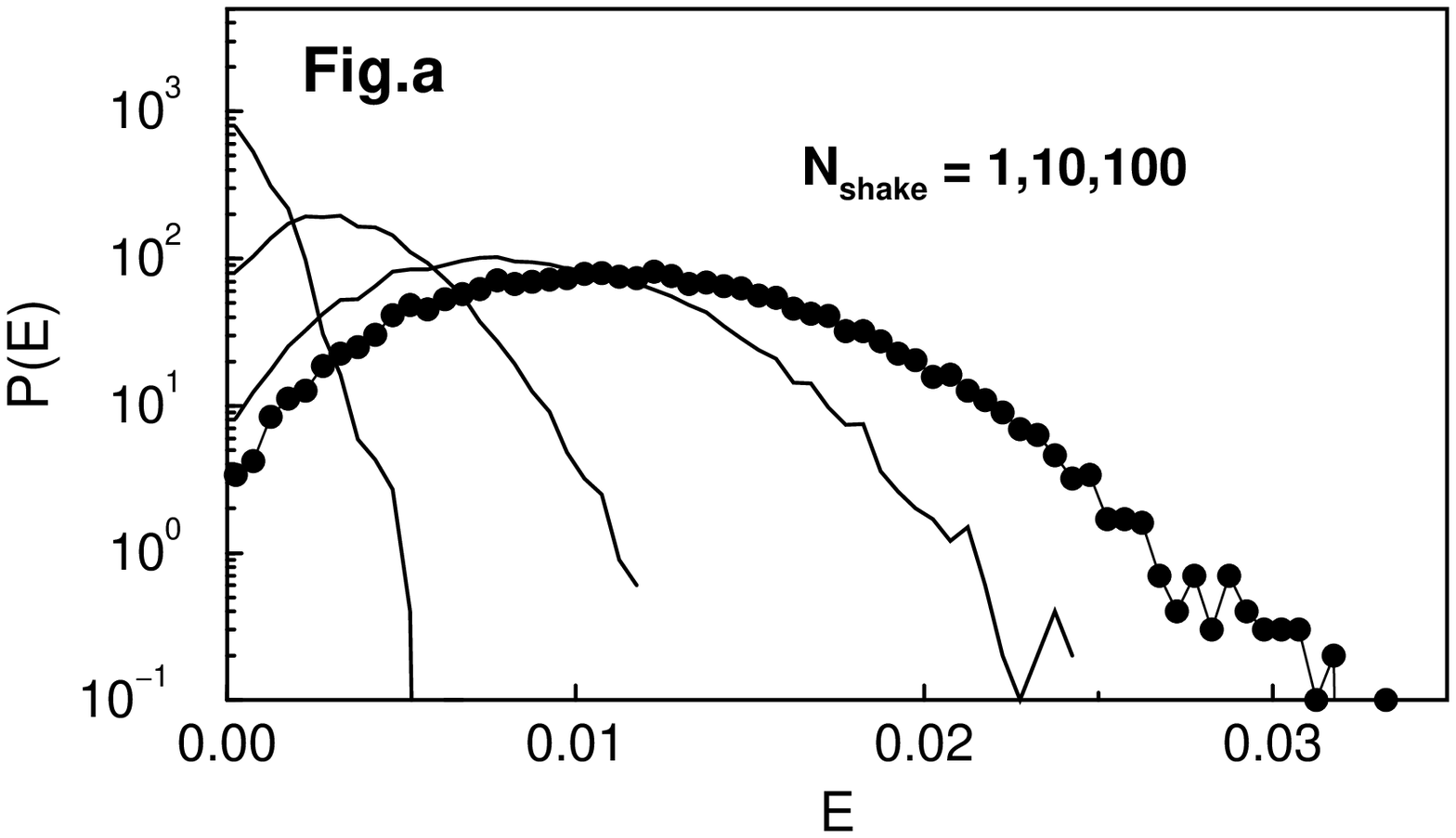} 
\epsfxsize=2.4in
\epsfbox{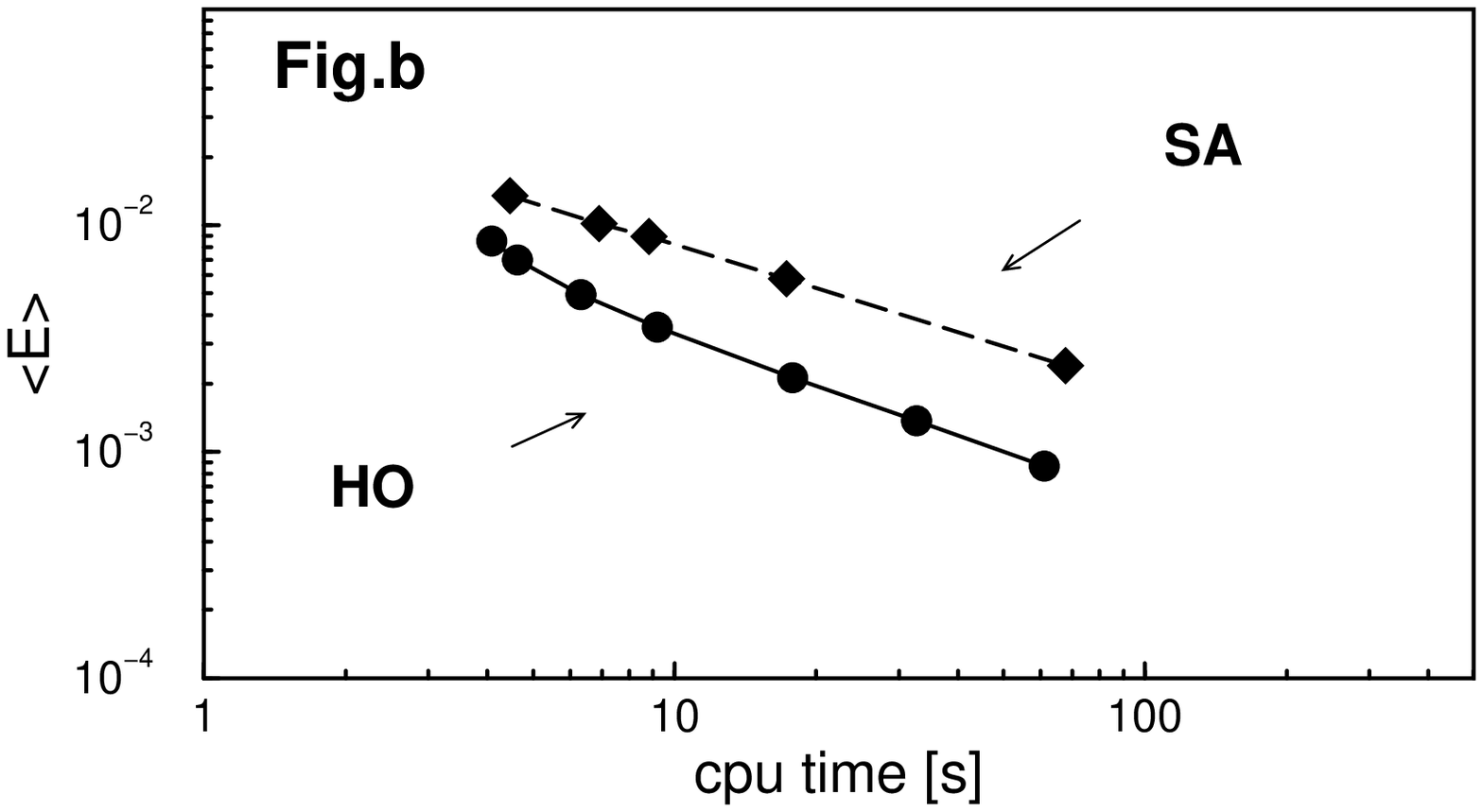}
\end{center}
\caption{\label{fig:sk_new} 
Fig.a: Probability density of finding a state of an SK model  
($N=1000$) with energy $E$  per spin above the ground state
by  a.c. demagnetization (symbols) 
and HO with 1, 10, and 100 shake ups  (solid lines).
Fig.b: Average energy  per spin measured from the ground state 
vs. running time for the SK model ($N=1000$).}
\end{figure}

\begin{figure}
\epsfxsize=2.7in
\begin{center}
\epsfbox{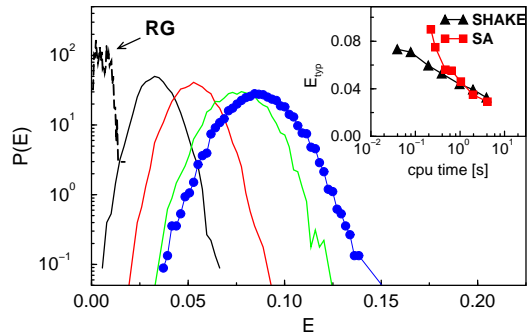}
\end{center}
\vspace*{0.05cm}
\caption{\label{fig:ea} 
Probability density of finding a state with energy $E$ 
per spin above the ground state 
for the three-dimensional ($L=10$) Edwards-Anderson model
by a.c. demagnetization
(symbols) and HO with 1, 10, and 100 shake ups (solid lines). 
The dashed line shows results of HO combined with RG. 
Inset: Typical energies  as a  
function of  running times for SA and HO with shake ups.
}
\end{figure}

The probability density $P(E)$ of finding a state with energy $E$
produced by  a.c. demagnetization
for the $N=1000$ SK and the  $10\times 10\times 
10$ EA models is shown by the symbols in Figs.~\ref{fig:sk_new}.a and 
\ref{fig:ea}. These measurements were performed for a single sample 
(instance).  The a.c. demagnetization
gives remarkable  improvement with respect to the quench, but 
in both cases the final state is definitely above the  ground state. 
We emphasize, however, that  a.c. demagnetization
runs 3-10 times faster than  SA  (see Figs.~\ref{fig:sk_new}.b 
and \ref{fig:ea}), 
and can be a viable alternative  in  applications where a large number 
of optimizations is required, each of them on relatively smaller systems.

As we mentioned earlier, unlike  SA, increasing the 
factor $\gamma$ closer to one, we only find a
limited effect on the quality of solutions obtained 
by a.c. demagnetization. This is a serious problem,
and finding the cause of it is a potentially important 
subject for future research. We speculate, however,
that the answer is hidden in the random, but ``frozen''
correlations between our noise term ${\xi_i}$ and the 
generated low-lying states.
These correlations may prevent us
from reaching certain parts of the phase space. If so,
all we have to do is to introduce different ${\xi_i}$-s,
since they have different correlations. But we have to
do this without destroying the ``good'' correlations
of previous ${\xi_i}$-s.

Maybe the simplest way to  achieve this
is ``shaking up'' the system:
After finishing an a.c. demagnetization, $H$ being zero, we are
free to change ${\xi_i}$ to a different noise ``direction'' 
${\xi'_i}$, while the final spin configuration ${\sigma^0_i}$
remains stable. Starting from this state, we increase the field
in small steps to $H^m$, quenching 
the spins after each step as before.
From this point we start a new demagnetization by decreasing
the field to $-\gamma_1 H^m$, and turning back e.t.c..
At the and of the shake up we get a new state ${\sigma^1_i}$, 
and $H$ is zero again, so we can choose a new ${\xi''_i}$, and
start another shake up. While subsequent shake ups with
appropriate $H^m$ improve the solution on average, it is
obviously helpful to get rid of ``bad fluctuations'', i.e.
in case ${\sigma^1_i}$ had higher energy than ${\sigma^0_i}$,
we would start the new shake up from the latter again.
It is this combination of a.c. demagnetization
and shake ups what we call hysteretic optimization (HO).

The exact choice of the   
shaking amplitude $H^m$
is best obtained by testing it with a few discrete values: too small
loops make hardly any change and too large field values take us back
to a.c. demagnetization results. Our practice shows that the best 
choice for $H^m$ is around the coercive field,
i.e. the field amplitude at which the magnetization reaches zero
if started from saturation 
($\sim 0.3$ on Fig. \ref{fig:loop}). 

Figs.~\ref{fig:sk_new}  and \ref{fig:ea} summarize our 
results of the shake-up HO calculations on the SK and EA models. 
Apparently, already the first few  shake-ups 
give substantial  improvement over  a.c. demagnetization. Increasing 
further the number of shake-ups is less and less effective, but we 
get systematically closer and closer to the ground state, 
similar to SA.  We would like to emphasize that applying  shake-ups 
after an a.c. demagnetization is straightforward, requires very little 
extra coding and computer time. A systematic comparison of 
a.c. demagnetization, shake-up HO, and SA is shown in 
 Figs.~\ref{fig:sk_new}.b and \ref{fig:ea}.

In the case of the short-range EA model  (inset of Figure~\ref{fig:ea})
shake-ups seem less effective, and for longer running times SA 
reaches the efficiency of HO. However, typical optimization 
problems are often closer to the long-range SK
model where no particular dimensionality is present 
(e.g. stock market) \cite{stock}.
In this case (Fig.~\ref{fig:sk_new}.b), HO appears to be better than SA
even for long running times.

We have to emphasize that, as well as SA
the HO presented in this paper should be considered             
as a  {\em building block}, and it can be efficiently combined
with other optimization methods,
such as genetic algorithms\cite{geneticalgorithms},
cluster renormalization group (RG) techniques \cite{rg},
or both \cite{martin}. To demonstrate this, we 
combined HO with the RG in the following way:
We generated clusters from 
$n(\sim 10-20)$ different low-energy configurations obtained by HO
and using the clusterization technique of Ref. \cite{martin}.
Using these cluster spins
simplifies the problem not just by decreasing the number of degrees
of freedom, but also by removing dynamical barriers: turning over
a bigger cluster through single spin flips might be energetically
expensive because of the strong couplings inside the cluster.
For the effective system of cluster spins HO can be applied
again to generate $n$ new (cluster) configurations and to define clusters
of clusters, and so on. 
We used this relatively large $n$\cite{one_d_remark}
on each RG level,
and consequently many RG steps,
to avoid the necessity of including a genetic algorithm.

It is worth noting that there is a
freedom in introducing the external field for the clusters,
since the ``magnetic moment'' $\mu$ of a cluster can be arbitrarily
chosen. The simplest choice is $\mu=1$ for all clusters, but
it might be useful to scale the local moments to insure that small
and large clusters can compete with each other.
The success of the RG approach for the EA model is due mostly to the 
short-range nature of the interactions (see Fig.~\ref{fig:ea}).
We find that this method is able to reach the 
ground state in a time comparable to   a recent
very efficient genetic algorithm developed specifically for this 
problem \cite{young,Karcsi}. 

Now we present a general scheme that makes it possible to apply 
the present algorithm to practically any optimization problem.
In an optimization problem one has  to minimize some cost function 
$W(P)$, which  is a mapping from some configuration 
space $\{P\}$ to  real numbers.  In order to apply HO to a system 
we need three important  ingredients: $(i)$ {\em Dynamics}. This is already 
sufficient to do SA,  however,  for the HO we need also a
$(ii)$  {\em distance}  $d(P,Q)$ over the 
configuration space, and
$(iii)$ two {\em reference states} $R_{\pm}$. 
These latter will be used to induce  
a random  external  field.  

We have a freedom  in chosing these ingredients. 
However, a `good' dynamics  must be such that  an elementary 
step connects configurations with approximately  equal cost functions,  
and its successive applications  connect all  
configurations.
Similarly, a 'good'  distance corresponding  to a given dynamics 
is such that distances  do not change much during a  single step.  

Having defined all these quantities we define the  HO  as the 
demagnetization of the Hamiltonian
\begin{equation}
{\cal H}(P) = W(P) 
+ \sum_{\alpha = \pm} \alpha H \Theta(\alpha H)
d(P,R_\alpha),
\label{eq:general} 
\end{equation}
whith $\Theta(x)$ being the step function. 
For the spin problems considered here 
the configurations are $P = \{\sigma_i\}$, the dynamics is 
single spin flip, the distance is  $d(\{\sigma_i\}, 
\{{\tilde \sigma}_i\}) = \sum_i |\sigma_i - {\tilde \sigma}_i|$, 
and the two reference states are $R_\pm = \{\pm \xi_1,..,\pm \xi_N\}$.
With these definitions Eq.~(\ref{eq:general}) reduces to 
Eq.~(\ref{eq:energy}), and the general HO becomes the one we applied to the 
spin problems. The basic role  of the  external field in 
Eq.~(\ref{eq:general}) is to force the system 
close to the states $R_\pm$ as $H\to\pm\infty$. 

\begin{figure}[tbh]
\epsfxsize=2.6in
\begin{center}
\epsfbox{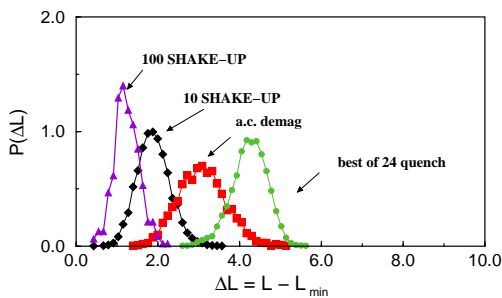}
\end{center}
\vspace*{0.05cm}
\caption{\label{fig:tsm} 
Distribution of length generated by HO for a two-dimensional 
TSP with $N=100$. For comparison, we also show the 
best of 24 quenches, taking the same time as a.c. demagnetization.}
\end{figure}

Now we demonstarate how this method can be applied for the TSP.
In this classical optimization problem the aim is to find 
the shortest path that goes through each of 
a given set of $i=1,..,N$ cities, visiting just once every 
one of them. In this problem the randomness is in the 
location ${\bf x}_i$ of the cities. 
The naive strategy of 
going always to the closest unvisited  city fails,
because typically there is a huge penalty for the last trip home ---
giving the major source of frustration. 
The  configuration space is just  given by all 
possible permutations $P$ of the cities, 
and the cost function is  simply the total length of the path, 
$W(P) = \sum_{i=1}^N | {\bf x}_{P(i)} - {\bf x}_{P(i+1)} |$.
As an elementary step we used the interchange of 
two cities \cite{remark}, and a distance
\begin{equation}
  d(P,Q)\equiv \sum_i^N | {\bf \Delta}_i(P) - {\bf \Delta}_i(Q)|,
\end{equation}
where ${\bf \Delta}_i(P)  = {\bf x}_{P(i+1)} - {\bf x}_{P(i)}$ 
denotes the vector connecting the cities $P(i)$ and $P(i+1)$. 
Our results for the traveling salesman are summarized in 
Fig.~\ref{fig:tsm}. We find a systematic improvement 
as the number of shake ups increases, and  get better and 
better results in this case too. 

There is one more aspect of HO we would like to emphasize.
SA has the nice property of self-consistently
telling us how the annealing schedule should be set: the specific
heat is proportional to the fluctuations, so one might want to
spend more time in the temperature regions of high specific heat.
A similar measure of accuracy for HO might be the {\it measured}
Preisach function of the system. This function
can be obtained numerically from first order reversal curves 
\cite{dellatorre},
and once at hand, it can guide the annealing of the loops.
We believe that this function has a lot of information about
the metastable states of the system, although admittedly future
research is needed to identify these features.

{\it Acknowledgements}:  This research has been supported  by  
NSF Grants No.  DMR99-81283, DMR97-14725 
and DMR99-85978 and Hungarian Grants No. OTKA F030041,  T029813, and 
29236.


\begin{references}
\bibitem{gs} We used the method of \cite{Karcsi} to obtain
the ground state for all models.
\bibitem{one_d_remark} This cluster RG can be solved exactly for
the one-dimensional EA spin glass.
\bibitem{remark}
While more complicated  dynamical steps \protect{\cite{kirkpatrick}}
can be more  efficient, here we chose a less efficient 
dynamics which allows one to more  clearly judge the performance
of HO. The ground state of the TSP was found using twist dynamics
\protect{\cite{kirkpatrick}}.
\end{references}
\end{document}